# A New Paradigm for Obtaining the Laws of Physics

*Edmund A. Di Marzio*

*BIO-POLY-PHASE*
*14205 Parkvale Road*
*Rockville, MD 20853*

**Abstract:**

The Lorentz transformation is derived without assuming the existence of Maxwell's equations, or that the speed of light is a constant, or even that light exists. This leads us logically to consider the existence of a field called primal-stuff which at every space-time point is traveling in every possible direction with the speed of light. All physical quantities are to be obtained from operations on the primal-stuff field. To familiarize ourselves with the implications of this new paradigm equations are developed for a world of time and only one spatial dimension ((1+1)-d). These non-linear equations lead to the existence of quantized particles, of light, and of gravity. The strong force, the electromagnetic force and the gravitational force are thereby unified. Conservation laws and symmetries are obtained. Spin space is introduced by assuming the primal-stuff field is a matrix. For a 3x3 matrix an eight-fold classification of particles arises and the equations describing their dynamical structures are obtained. The extreme simplicity of the equations suggests that they may have a (1+3) dimensional counterpart which is adequate to unify the four forces of our real world. A candidate (1+3)-d equation is proposed.

**(1) Introduction:**
Some time ago it was proposed that physical reality can be described by an equation for a field $\rho(x, \Omega)$ where $x$ is the space-time point $(t, x, y, z)$ and $\Omega$ is a direction at $x$. The $\rho(x, \Omega)$ field represents primal-stuff traveling always with the speed of light in the direction $\Omega$, and all $\Omega$ are allowed simultaneously at each point $x$. See Fig. 1. Thus, there are an infinity of equations at each $x$; one for each $\Omega$. It was suggested that $\rho(x, \Omega)$ is a null vector density.

If we imagine ourselves to be riding along one of the two-fold infinity of directions $\Omega$ at $x$, with the speed of light then, by the principle of contiguous action, the only way $\rho(x, \Omega)$ can change is by interacting with other $\rho(x, \Omega_1)$ at the same space-time point $x$. Thus, the equation for $\rho(x, \Omega)$ will have the general form

$$D\rho(x, \Omega)/Dt = \int f(\rho(x, \Omega), \rho(x, \Omega_1)) d\Omega_1 \qquad (1)$$

where $D\rho(x, \Omega)/Dt$ is the co-moving derivative. As we shall see, $\rho(x, \Omega)$ is both a null vector density and a hypercomplex number (a matrix). In previous works we proposed functional forms for the RHS of Eq. 1 when there is one spatial dimension plus time (a (1+1) dimensional world).[1,2] Equations were suggested for the cases of $\rho(x, \Omega)$ being a complex number[1] and for $\rho(x, \Omega)$ being a 2x2 matrix of complex numbers[2]. We also explored the case where $\rho(x, \Omega)$ is a matrix in our real world of (1+3) dimensions[3].

We defined a particle as the existence of a primal-stuff field that remains in place even though every part of the field consists of primal-stuff traveling always with the speed of light. In the (1+1)-d world we discovered the existence of such particles. The particles were quantized by virtue of the non-linearity of the equations, and because the equations were Lorentz covariant the particles could be made to move with any constant speed less than c. Further, the equations admitted of two-particle solutions and n-particle solutions which displayed creation and annihilation of particles, always in pairs. Further, the general solution of the equations was obtained. We emphasize that the existence of particles was not assumed, rather their existence is a natural consequence of the equations as are also their properties. In light of the above we see that we have in (1+1)-d the analogue of the strong force of our real world of (1+3)-d.

Our previous treatments were preoccupied with the amazing fact that one can have stationary particles in space even though the primal-stuff field that comprised them is always traveling with the speed of light. Our treatment of electromagnetism was quite heuristic and the generalization from Lorentz covariance to general covariance was sketchy[3]. Also we had not derived the implications of replacing primal-stuff by a matrix.

In this paper we shall revisit the one dimensional case. We will show rigorously that that the equations imply electromagnetism. We will generalize from Lorentz covariance to general covariance so that an attempt to relate **ρ(x, Ω)** to the metric can be made. In this way gravity in our (1+1) dimensional world can be investigated much as Einstein did in the real world. The strong force, electromagnetism and gravity are thereby unified. Additionally we shall investigate the properties of our (1+1)-d world when **ρ(x, Ω)** is a 3x3 matrix. This is useful because it allows us to classify particles in an eight-fold way.

Working in (1+1) dimensions allows us to familiarize ourselves with techniques that will carry over to the real world of (1+3) dimensions. We shall not try to develop details of the three dimensional case in this paper but it is obvious that we have an eye towards that problem. Occasionally we shall comment on the (1+3) dimensional case. In **Appendix A** we suggest a candidate equation for (1+3)-d.

We are striving for complete comprehensibility. For this reason we will not simply present the equations in (1+1) dimensions but will instead base them on more primal principles, to the maximum extent possible. As we show below the Lorentz transformation (LT) is such a principle since it can be derived assuming only that meter sticks and clocks exist and that they can be synchronized. Remarkably, the LT is derived without assuming that the speed of light is a constant, or assuming the existence of Maxwell's equations, or even that light exists[4]. The LT is then even more primal than heretofore assumed. A concise outline of the derivation of the LT contained in Reference 1 follows.[1] Our derivation will lead us to added insights.

**(2) Derivation of the Lorentz Transformation for the Real World of Three Dimensions and Time:**
Let starred and unstarred coordinate systems move with constant velocity v along a line containing the x and x* axes which are pointing towards each other. The symmetry of this situation allows us to write

$$x^* = \alpha x + \beta t \qquad x = \alpha x^* + \beta t^*$$
$$t^* = \delta x + \varepsilon t \qquad t = \delta x^* + \varepsilon t^* \tag{2}$$

$$y^* = y, \; z^* = z$$

In matrix form we can write Eqs. 2 as

$$\begin{vmatrix} x^* \\ t^* \end{vmatrix} = \begin{vmatrix} \alpha & \beta \\ \delta & \varepsilon \end{vmatrix} \begin{vmatrix} x \\ t \end{vmatrix} \qquad \begin{vmatrix} x \\ t \end{vmatrix} = \begin{vmatrix} \alpha & \beta \\ \delta & \varepsilon \end{vmatrix} \begin{vmatrix} x^* \\ t^* \end{vmatrix} \tag{3}$$

Symbolically

$$\mathbf{x^*} = \mathbf{Ax}, \qquad \mathbf{x} = \mathbf{Ax^*} \tag{4}$$

Thus, $\mathbf{A}^2 = \mathbf{1}$, or

$$\begin{vmatrix} \alpha & \beta \\ \delta & \varepsilon \end{vmatrix}^2 = \begin{vmatrix} 1 & 0 \\ 0 & 1 \end{vmatrix} \tag{5}$$

The $\alpha$, $\beta$, $\delta$, $\varepsilon$, are each independent of the coordinates because of homogeneity in space and time. From Eqs. 5 we obtain 4 equations, which allow us to express 2 of the constants in terms of the others.

$$\mathbf{A} = \begin{vmatrix} \pm(1-\beta\delta)^{1/2} & \beta \\ \delta & \mp(1-\beta\delta)^{1/2} \end{vmatrix} \tag{6}$$

The constant $\delta$ can be removed by observing that the relative velocity of the two coordinate systems is v. Thus (x*=0, t*) implies (x=vt) which gives

$$\mathbf{A} = \begin{vmatrix} \mp \beta/v & \beta \\ (1/\beta - \beta/v^2) & \pm \beta/v \end{vmatrix} \tag{7}$$

To obtain $\beta$ we first make a transformation which points the positive axes in the same direction and then use the transitive rule. Symbolically, we have

$$\mathbf{x^{**}} = \mathbf{A_3 x}, \quad \mathbf{x^{**}} = \mathbf{A_2 x^*}, \quad \mathbf{x^*} = \mathbf{A_1 x}; \qquad \mathbf{A_3} = \mathbf{A_2 A_1} \tag{8}$$

Or, in detail

$$\begin{vmatrix} \pm \beta_3/v_3 & -\beta_3 \\ -(1/\beta_3 - \beta_3/v_3^2) & \pm \beta_3/v_3 \end{vmatrix} = \begin{vmatrix} \pm \beta_2/v_2 & -\beta_2 \\ (1/\beta_2 - \beta_2/v_2^2) & \pm \beta_2/v_2 \end{vmatrix} \begin{vmatrix} \pm \beta_1/v_1 & -\beta_1 \\ (1/\beta_1 - \beta_1/v_1^2) & \pm \beta_1/v_1 \end{vmatrix}$$

The equations from the (1,1) and (2,2) positions of the matrix result in

$$\frac{1}{\beta_1^2} - \frac{1}{v_1^2} = \frac{1}{\beta_2^2} - \frac{1}{v_2^2} = -\frac{1}{c^2} \tag{9}$$

Where c is, at this point, an unknown constant with the dimension of velocity. Placing the values of β, (β=v(1-(v/c)² )$^{-1/2}$ ) into the above expressions for the A's of Eq. 8 we obtain the Lorentz transformation

$$\begin{vmatrix} x^* \\ t^* \end{vmatrix} = \gamma \begin{vmatrix} 1 & -v \\ -v/c^2 & 1 \end{vmatrix} \begin{vmatrix} x \\ t \end{vmatrix}, \quad \begin{vmatrix} x \\ t \end{vmatrix} = \gamma \begin{vmatrix} 1 & v \\ v/c^2 & 1 \end{vmatrix} \begin{vmatrix} x^* \\ t^* \end{vmatrix} \tag{10}$$

where γ=(1-(v/c)²)$^{-1/2}$ . The equations from the (1,2) and (2,1) positions of the matrix each separately result in the addition law for the velocities.

$$\gamma_3 v_3 = \gamma_2 \gamma_1 (v_2 + v_1) \tag{11}$$

The usual form for addition of velocities ( $v_3 = (v_1 + v_2)/(1 + v_1 v_2 / c^2)$ ) which is easily derived from Eq. 10, when substituted into the left hand side of Eq.11 results in an identity.

From the symmetry of our initial construction we also see that y*=y and z*=z. Thus the Lorentz transformation is valid in a world of three spatial dimensions plus time. We have not proved that it is valid globally, but we have proved that is valid locally. That is to say that at every point in space time we could have made the above arguments; but to imagine that the LT would be valid at large distances would be an unwarranted assumption. Further, by substitution into Eqs. 10 it is a simple matter to show that in every local coordinate system the metric reads

$$-c^2(dt)^2 + (dx)^2 + (dy)^2 + (dz)^2 = -c^2(dt^*)^2 + (dx^*)^2 + (dy^*)^2 + (dz^*)^2 = ds^2 \tag{12}$$

What have we assumed in our derivation? Only that meter sticks and clocks exist and that the clocks can be synchronized. One classical way to synchronize clocks is to move them slowly from one place to another. Another way is to use the sound waves traveling through the meter sticks. So our derivation really is fundamental. The LT transformation is a principle of which we are absolutely certain.

We now seek to determine the significance of c. The first thing to observe is that according to Eqs. 10 a particle moving with velocity c in one coordinate system moves with velocity c in every coordinate system. To prove this one simply replaces x* by Vt* in Eqs. 10 and takes the limit as V approaches c. The relation x = ct results. Alternatively one can use the formula for the addition of velocities directly. We know from experiment that this is true for the speed of light in vacuum so we can write c= 2.998x10$^8$m/sec. But there is more to the meaning of c than just its numerical value. Just think of it. We have with pure thought derived a set of equations which we know has revealed an important fact of reality; the existence of a universal velocity. Additionally, once the LT is obtained many of the conclusions of special relativity follow, with minimal added assumptions, including the famous relation of Einstein[5] E=mc$^2$. These are all consequences of the Lorentz Transformation.

There is an important philosophical difference between our derivation of the LT and previous derivations. In our derivation we first used pure thought to derive a mathematical relation and then the experimental fact that the speed of light is a constant validates the notion that pure thought is adequate to the task of discovering the most basic of realities. In effect we have replaced Descartes **I think therefore I am** by the **Lorentz Transformation** as the prime epistemological reality.[6] We have shown that the view that many geometries are a priori possible but the experimental fact that c=constant picks out the LT is a false view. Rather we have shown that the (local) LT is the only logical possibility.

Kant's view that the human mind can sense only the appearance of things and not their basic reality is also shown to be false.[7] Meter sticks and clocks are easily apprehended by the senses and they lead to immediate deep knowledge of ultimate realities such as the constancy of the speed of light.

Thus the experimental discovery that the speed of light is a constant becomes a great confidence builder. It suggests that pure thought is adequate to the task of deriving the laws of physics and establishes the LT as the prime epistemological reality upon which to build the edifice of physical law.

**(3) Statement of the Basic Paradigm:**
We now seek to derive the laws of physics by proceeding logically from the LT as base since, as noted above, the LT is even more fundamental than E=mc$^2$. At the very least we know that the quantity c must enter into a significant statement concerning the structure of atoms comprising the meter sticks and clocks, since their existence is the only thing we have assumed in our derivation. So we ask, can we devise a theory in which our derivation of the LT suggests a new possibility?

Consider the following. We can seek to devise a theory in which only one speed occurs. We can seek to devise a theory in which only two speeds occur? …We can seek to devise a theory in which only n speeds occur… It is immediately obvious that only the one-speed case is a relativistically covariant statement. As long as n is finite, no matter how large, only the case n=1 is relativistically covariant. However if within every small increment δ of line element, no matter how small, between -c and +c, there is an infinity of velocities then my thinking breaks down.

Ockham's razor dictates that in a logical development one first chooses and investigates the option that has the greatest simplicity. So we will assume as a new paradigm only one speed since this is obviously much simpler than the continuum of speeds cases.

But we are immediately faced with an apparent contradiction because we know from experience that particles can move with speeds less than the speed of light. This apparent contradiction can be easily resolved by assuming that particles are extended objects describable by a "primal-stuff" field ρ(t, x, y ,z; **Ω**) in which at each point (t, x, y, z) there is stuff traveling in each possible direction **Ω,** always with speed c (see Fig. 1). Primal-stuff traveling in direction **Ω** at the space-time point **x** can change only by being scattered by stuff **Ω$_1$** at the same point. Thus, the reason for the form of Eq. 1 has now been explained. Of course we have not yet proved that the RHS of Eq. 1 allows the existence of particles that remain stationary in space even though they are composed entirely of stuff that is always traveling with the speed of light. This is done in **Section 4**.

The view that the only existing physical reality is primal-stuff allows us to make a second significant discovery; namely that when applied to the world of physics, Weyl geometry[8] and Riemannian geometry are one and the same. See **Appendix A**.

The primal-stuff field is a mathematical object and we will assume that it is a vector density. Spinors and twisters also might work but we have not investigated them. We choose a vector density because we want its space integral to be a geometrical object.[9] For the special case of Lorentz covariance a vector density is equivalent to a vector since the Jacobian is +1. We choose it to be a null vector density because then it represents primal-stuff always traveling with the speed of light. Once we have assumed a vector density the whole bag of tricks of the absolute differential calculus is available to us.[10]

We now proceed to derive the equations for a toy world of one dimension plus time. Ultimately, we will obtain a set of equations for $\rho(t, x, y, z; \Omega)$ that are covariant under arbitrary transformations, (we do this in **Section 4.6** for the (1+1) dimensional case**)** and we seek another set of equations that relate the first set of equations to the metric (see **Section 4.7).** This will give us a perspective on how to attack the problems of our real world of (1+3) dimensions.

**(3.1) Implementation of the Paradigm for (1+1) Dimensions:**
In the remainder of this paper we first describe the simplest set of (1+1) dimensional equations subject to the LT globally (Eqs. 15). We then display one particle, two particle and, n particle solutions. Quantization of the particles arises from the nonlinear character of the equations. We obtain the general solution of this pair of first-order non-linear partial differential equations (Eqs. 23). We offer a candidate for the short-range force law between two particles (Eq.22) and outline a procedure for obtaining the long-range force.

To prove that our basic Eqs. 15 imply a (1+1)-d electromagnetism we first demonstrate that the structure of the microscopic Maxwell equations ((1+3)-d ) is completely contained in the Lorentz condition alone, (Eq. 25A), which is a conservation law on the electromagnetic four vector.[11] The analogous conservation law (Eq. 26) for (1+1)-d is obtained and (1+1)-d electrodynamics is thereby derived.

Primal-stuff obeys the same transformation law as the relativistic Doppler effect (Eq. 13). This leads us to wonder if primal stuff can be interpreted as energy. For the one particle solution we display what seems to be the energy-momentum relation, Eq. 29, and derive $E=mc^2$.

The equations symmetries and group properties are obtained in **Section 4.5**.

Having obtained the source of the strong force and the electromagnetic force in our (1+1) dimensional world we then express the equations in generally covariant form so that gravitation can be investigated via the route of relating the metric to matter. Eqs. 35 and 36 are two possible generalizations.

We then introduce "spin" by making the fundamental stuff vector a matrix rather than just a complex number. For the case of a 3x3 matrix we obtain an eight fold way which allows us to obtain a particle Zoo (See the set 47). Conservation laws and symmetries of these equations are discussed.

In **Appendix A** we show that, given the paradigm that only the primal-stuff field ρ(x; Ω) exists it alone can be used to transfer length from point to point, Weyl geometry then becomes identical to Riemannian geometry.

The relation of our equations to variational principles is discussed in **Appendix B**.

In **Appendix C** we display the (1+3)-d equation which is easily derived. It seems to be unique.

In **Appendix D** we observe that if our equations are fundamental then they must underlie quantum mechanics. Some notes on the problem of relating our paradigm to quantum mechanics are made.

**(4) Equations for a One-Dimensional World:**
The transformation law for primal-stuff is easily obtained from the LT as

$$(\rho_R^0)' = \gamma(\rho_R^0 - \rho_R^1(v/c)) \qquad (\rho_L^0)' = \gamma(\rho_L^0 - \rho_L^1(v/c))$$
$$(\rho_R^1)' = \gamma(-\rho_R^0(v/c) + \rho_R^1) \qquad (\rho_L^1)' = \gamma(-\rho_L^0(v/c) + \rho_L^1)$$
(13)

Using $\rho_R^0 = \rho_R^1$ and $\rho_L^0 = -\rho_L^1$, we see first, that null vector densities transform as null vector densities, and second, that the law of transformation is identical to the Doppler law. This is true also for (1+3)-d.

Consider the following equations using $\rho_R = \rho_R^0 = \rho_R^1$ and $\rho_L = \rho_L^0 = -\rho_L^1$.

$$\frac{\partial \rho_R(t,x)}{\partial t} + c\frac{\partial \rho_R(t,x)}{\partial x} = \rho_R(t,x)\rho_L(t.x) \qquad (14A)$$

$$\frac{\partial \rho_L(t,x)}{\partial t} - c\frac{\partial \rho_L(t,x)}{\partial x} = -\rho_R(t,x)\rho_L(t.x) \qquad (14B)$$

ρR(t,x) represents stuff traveling to the right with speed c and ρL(t,x) represents stuff traveling to the left with speed c. The left hand sides are commoving derivatives, in agreement with Eq. 1. If we imagine ourselves to be riding along the waves then the only way that ρL(t,x) ( ρR(t,x)) can change is by being scattered by ρR(t,x) ( ρL(t,x)). Thus, our equations obey the principle of contiguous action.

If we divide by c and rescale the rhos we obtain

$$\frac{\partial \rho_R(ct,x)}{\partial ct} + \frac{\partial \rho_R(ct,x)}{\partial x} = \rho_R(ct,x)\rho_L(ct.x) \qquad (15A)$$

$$\frac{\partial \rho_L(ct,x)}{\partial ct} - \frac{\partial \rho_L(ct,x)}{\partial x} = -\rho_R(ct,x)\rho_L(ct.x) \qquad (15B)$$

In this form we see that the LHS' are also divergences. A point of notation: some times for the purpose of clarity of presentation we will set c=1, sometimes ct=$x^0$, x=$x^1$.

Because the stuff is traveling with speed c they are null vectors and $\rho_R \circ \rho_R = 0$, $\rho_L \circ \rho_L = 0$.

In component form we have

$$\rho_R = (\rho_R{}^0, \rho_R{}^1) = (\rho_R{}^0, \rho_R{}^0) = (\rho_R, \rho_R), \quad \rho_L = (\rho_L{}^0, \rho_L{}^1) = (\rho_L{}^0, -\rho_L{}^0) = (\rho_L, -\rho_L) \quad (16)$$

These statements are valid for any coordinate system because the vectors are null vectors (they are always traveling with the speed of light) in any coordinate system. The rhos give the amount of stuff but the speed of stuff is always c. The RHS of the equations are dot products and the LHS of the equations are divergences. So the equations are Lorentz convariant. This can be verified directly by transforming Eqs. 15 by use of the equations

$$\rho_R^* = \gamma(1+v/c)\rho_R$$
$$\rho_L^* = \gamma(1-v/c)\rho_L \qquad (17)$$

$$x^* = \gamma(x+vt)$$
$$t^* = \gamma(t+vx/c^2) \qquad (18)$$

**(4.1) The General Solution Leads to Many Particles:**
One sees immediately that a solution of Eqs. 15 corresponding to a stationary particle exists. Consider

$$\rho_R = \rho_L = -1/(x-x_0) \qquad (19)$$

The stuff traveling to the right scatters stuff traveling to the left (and conversely) in such a way as to maintain the particle in existence at a given place. Notice that these particles are quantized. $-k/(x-x_0)$ is a solution only for k=1. Note that Eqs. 15 describe how any initial condition evolves in time and space. Only for $\rho_R = \rho_L = -1/(x-x_0)$ does the evolving solution remain $\rho_R = \rho_L = -1/(x-x_0)$. For $\rho_R = \rho_L = -k/(x-x_0)$, $k \neq 1$ there is evolution to something else. An evolving solution of Eqs. 15 that remains in place for a long time is called a particle, or particles.

Note well that if $\rho_R$ is negative in value this does not mean that it is traveling to the left. It means only that the numerical value of the stuff traveling to the right, $\rho_R$, is negative. If $\rho_R$ were imaginary it would mean only that the stuff traveling to the right were imaginary in value. We see no reason, at this time, to limit ourselves to real numbers.

Thus the equations allow for the existence of particles. If we perform a boost on Eq. 19 we obtain

$$\rho_R = -\frac{(1+v/c)}{x-vt+x_0(1-(v/c)^2)^{1/2}} \qquad (20)$$

$$\rho_L = -\frac{(1-v/c)}{x - vt + x_0(1-(v/c)^2)^{1/2}}$$

It is easily verified by substitution into Eqs. 15 that Eqs. 20 are solutions..
  A two particle solution is

$$\rho_L = -\frac{2(x+t)}{x^2 + t^2 - A^2} \tag{21}$$

$$\rho_R = -\frac{2(x-t)}{x^2 + t^2 - A^2}$$

The centers of the two particles are located at $x = \pm\sqrt{A^2 - t^2}$ Only for (-A<t<A) can we locate each of the two particles by their $1/(x-x_0)$ like infinities (the places where the denominator is zero). Outside this range there is no infinity (the x's are imaginary). That is to say that the equation $x^2 + t^2 - A^2 = 0$ results in x being imaginary. Notice that the centers of the particles as soon as they are created separate from each other with infinite speed and then they travel towards each other with speed $\frac{dx}{dt} = \pm 2t/\sqrt{A^2 - t^2}$ and they annihilate themselves with the stuff traveling away with the speed of light. Whether they are particles or light is determined by whether the quadratic term in the denominator has real solutions or imaginary solutions. This seems to be particles being created and annihilated in pairs.
  We can now derive an analogue of Newton's law by writing

$$\frac{dx}{dt} = \mp 2t/\sqrt{A^2 - t^2}$$

$$\frac{d^2x}{dt^2} = \mp 2A^2/x^3 \equiv F/m \tag{22}$$

The separation between the two particles is 2x. The force is inversely proportional to $x^3$ and proportional to $A^2$. At this point we don't know if we are describing the analogue of the strong force, charge interaction or even mass interaction. An alternative approach is to place two particles each of which is described by Eq. 19 a great distance apart and then see how they accelerate. Since the sum of two solutions is not a solution we know that there will be acceleration. This problem is easily solved numerically on a computer. The results can be viewed as the solution to the equation $F/m = d^2x/dt^2$. This approach gives Newton's law in the weak interaction limit. The question of the additivity of forces can also be investigated in this way. Another question amenable to computer calculations is under what conditions the particles are stable to perturbations.
  One verifies by inspection that the general solution of Eqs.15 is

$$\rho_R = -2\frac{df(\nu)}{d\nu}[f(\nu) + g(\mu)]^{-1} \tag{23}$$

$$\rho_L = -2\frac{dg(\mu)}{d\mu}[f(\nu) + g(\mu)]^{-1}$$

$$\nu = (x - t) \qquad \mu = (x + t)$$

where g and f are arbitrary functions.

One can obtain a plethora of particles from Eqs. 23 by choosing g and f to be polynomials in (x+ct) and (x-ct). One can then view f(ν) +g(μ) as a polynomial in x. From the factor theorem for polynomials we write.

$$f(x - t) + g(x + t) = \prod_n (x - x_n(t)) \tag{24}$$

Where the x(t) that are imaginary occur in conjugate pairs. The particle centers are located wherever the denominator of Eqs. 23 equals zero.
. It seems that particles are nothing more than bottled up light. But instead of light being reflected by the interior of the surface of the bottle it is reflected according to the scattering law (dot product) described by the RHS of Eqs. 15.

Using the "four forces of matter" classification we have our strong force. In our world of (1+1) dimensions particles exist, they are quantized, and they can interact. By use of Eqs. 23 the stability of these particles can be investigated.[2]

**4.2 Electromagnetism in (1+1)-d:**
Jackson observes[12] that in (1+3)-d the microscopic Maxwell equations are "equivalent in every way" to the following equations.

$$\frac{\partial A^\mu}{\partial x^\mu} = 0, \qquad \mu = 0, 1, 2, 3 \tag{25A}$$

$$\Box A^\mu = \frac{\partial^2 A^0}{\partial (ct)^2} - \frac{\partial^2 A^1}{\partial x^2} - \frac{\partial^2 A^2}{\partial y^2} - \frac{\partial^2 A^3}{\partial z^2} = \frac{4\pi}{c} j^\mu, \tag{25B}$$

Here $A^\mu$ is the electromagnetic four vector and $j^\mu$ is the charge-current vector. $\Box$ is the d'Alembertian. Since derivatives commute one easily derives the conservation law for charge-current from Eqs 15A and 15B.

$$\frac{\partial j^\mu}{\partial x^\mu} = 0, \tag{25C}$$

In a previous paper[11] we showed that the microscopic Maxwell equations are equivalent to the Lorentz condition (Eq. 25A) alone. The basic reason for this is that the d'Alembertian is an invariant operator so that the d'Alembertian operating on any vector gives a vector. Thus, Eqs. 25B and 25C are implied by Eq. 25A alone.

Notice then, that the microscopic Maxwell equations are essentially equivalent to a conservation law alone. Conservation laws by their very nature imply underlying basic equations of which they are the conservation laws. The laws of mechanics are a clear example. The momentum and energy conservation laws imply the existence of underlying laws which are in fact Newton's laws of motion, and of course the conservation laws are logical consequences of Newton's laws. For the case of the (1+3)-d conservation law of electromagnetism we do not yet know the underlying equations of which they are a conservation law. But we are aiming to discover them.

However for (1+1)-d we do have the underlying basic equations so all we need do is identify the conservation law arising from Eqs. 15. To obtain the (1+1)-d analogue of the Maxwell equations we observe that if we sum Eqs. 15A and 15B we obtain a conservation law.

$$\frac{\partial(\rho_R + \rho_L)}{\partial ct} + \frac{\partial(\rho_R - \rho_L)}{\partial x} = 0 \tag{26}$$

Or equivalently

$$\frac{\partial A^\mu}{\partial x^\mu} = 0, \qquad \mu = 0, 1 \tag{26A}$$

$$\Box A^\mu = \frac{\partial^2 A^0}{\partial(ct)^2} - \frac{\partial^2 A^1}{\partial x^2} = \frac{4\pi}{c} j^\mu, \qquad \mu = 0, 1 \tag{26B}$$

where $\rho_R + \rho_L$ is the zeroth component and $\rho_R - \rho_L$ is the 1th component of the sum $\boldsymbol{\rho}_R + \boldsymbol{\rho}_L$ See Eq. 16. Eqs. 26A and 26B are a (1+1) dimensional version of the equations of electricity and magnetism. The Lorentz condition corresponds to Eq. 26A with the electromagnetic vector potential $\mathbf{A} = \boldsymbol{\rho}_R + \boldsymbol{\rho}_L$, and Eqs. 26B correspond to the multiplication of the electromagnetic potential by the invariant d'Alembertian.

We conclude that Electromagnetism is contained in the basic equations, Eqs. 15A and 15B. Thus we have unified two of the four forces of nature in (1+1)-d.

In Reference 3 we obtained the (1+3) dimensional analogue of Eq. 25A where the sum of null vectors, $\mathbf{Q} = \int \boldsymbol{\rho}(t,x,y,z; \boldsymbol{\Omega}) d\boldsymbol{\Omega}$ was identified with the electromagnetic 4 vector $\mathbf{A}$ $(\mathbf{Q}=\mathbf{A})$. However, one does not evaluate $\mathbf{A}$ by knowing $\mathbf{J}$ as one does in the classical view of electrodynamics, rather one uses the basic equations to evaluate $\mathbf{A}$ and then the conserved current $\mathbf{J}$ is determined by operating on $\mathbf{A}$ with the invariant d'Alembertian.

### 4.2.1 An Infinity of Conservation Laws:
Notice that we can operate on the vector of Eq. 25C with the d'Alembertian to obtain a new conservation law. But we could have done this repeatedly and as often as desired. This means that there are an infinity of conservation laws. In a somewhat different context Miura suggests that when an equation displays an infinite number of conservation laws one should expect the existence of solitons.[13] Notice also that there is a family resemblance of Eqs. 14 to the equations of soliton theory.[14]

### 4.3 Doppler's Law and Energy-Momentum:

Eqs. 13 show that primal-stuff transforms identically as Doppler's law. This strongly suggests, but does not prove, that primal stuff is energy. To prove this we will use essentially the same argument as used by Einstein in deriving $E=mc^2$. [5]

Let us examine the sum of the two null vectors which define the particle $\rho_R = \rho_L = -1/x$. In component form we have

$$(\rho_R^0 + \rho_L^0, \rho_R^1 + \rho_L^1) = (\rho_R^0 + \rho_L^0, \rho_R^0 - \rho_L^0) = (\rho_R + \rho_L, \rho_R - \rho_L) = (\rho_R + \rho_L, 0) \quad (27)$$

where the first equality arises because the vectors are null vectors (the sum is not), and our second equality arises from our simplified notation. The value of the invariant dot product is $-(2\rho_R)(2\rho_L)$ and since for our particle $\rho_R = \rho_L$ we set $2\rho_R = E$ Let us now move towards our particle with a velocity v. We have

$$(\rho_R^{*0} + \rho_L^{*0}, \rho_R^{*1} + \rho_L^{*1}) = (\rho_R^{*0} + \rho_L^{*0}, \rho_R^{*0} - \rho_L^{*0}) = (\rho_R^* + \rho_L^*, \rho_R^* - \rho_L^*) \quad (28)$$

Substituting for the rhos we obtain using Eqs. 17

$$(\gamma E, \gamma(v/c)E) = (E^*, \gamma c v (E/c^2)) \quad (29)$$

Now expanding $\gamma$ we obtain

$$E^* = \gamma E = E + \frac{1}{2}v^2(E/c^2) + ... \quad (30)$$

If we accept the experimental fact that the energy of a moving system equals that of the stationary system plus kinetic energy then $E/c^2$ must be equal to mass and E equal to the rest energy. Eq 29 becomes the energy momentum vector. Notice that the dot product of the energy-momentum vector with itself is $E^2$. Thus the existence of mass resides in the fact that the sum of null vectors is not itself a null vector. This proof is for just a small piece of the field, but it is easy to show that it holds also when we integrate over x. It is only required that

$$\int \rho_R dx = \int \rho_L dx \quad (31)$$

Although the above proof is highly suggestive that in (1+3)-d primal-stuff is energy we do not claim that it can be carried over to (1+3) dimensions unchanged for two reasons. First, other vectors can be formed from the rhos and their derivatives and they have to be examined. Second, when we generalize to a 3x3 matrix in **Section 4.10** we will obtain 9 conservation laws and we do not know what combination of those conservation laws might correspond to energy.

If $\rho_R$ and $\rho_L$ are small the RHS's of Eqs. 15 can be set to zero and we obtain

$\rho_R$ = g(x-ct)

$\rho_L$ = f(x+ct) \quad (32)

where g and f are arbitrary but small functions. If we now transform to a new coordinate system moving with velocity v, we obtain by use of Eqs, 17

$$\rho_R^* = \gamma(1+v/c)g(\gamma(1+v/c)(x^* - ct^*))$$
$$\rho_L^* = \gamma(1-v/c)f(\gamma(1-v/c)(x^* + ct^*))$$
(33)

If we initially had a sine wave function for g(x-ct) (g(x-ct) = Sin($\omega_0$ (x-ct)), on transforming we obtain g = Sin($\omega_0$ $\gamma$(1+v/c)(x*- ct*)) so that the relativistic Doppler effect $\omega_1$ = $\omega_0$ $\gamma$(1±v/c), is derived. But Eqs. 32 show that stuff itself follows the same law, so that it makes sense, for the present at least, to assign the concept of energy to the primal-stuff field.

**(4.4) Equations Equivalent to a minimization Principle:**
Eqs. 15 can be derived through a variational principle. See **Appendix B**.

**(4.5) Group Structure of the Equations:**
The groups are important because they reveal symmetries and because given a solution then we can generate other solutions (families of particles) by applying the group operations. They are:

**The Scaling Group:**
If $\rho(t,x)$ is a solution then $k\rho(kt,kx)$ is a solution.

**The Four Group:**

$$\rho_R = \rho_R(t,x) \quad ; \qquad \rho_L = \rho_L(t,x) \qquad (34)$$

$$\rho_R = -\rho_R(-t,-x) \quad ; \qquad \rho_L = -\rho_L(-t,-x)$$

$$\rho_R(t.x) = +\rho_L(-t.x) \quad ; \qquad \rho_L = +\rho_R(-t,x)$$

$$\rho_R = -\rho_L(t,-x) \quad ; \qquad \rho_L = -\rho_R(t,-x)$$

This is the Abelian four group.[15] Notice that parity is not a symmetry operation; nor is time reversal. The basic equations (Eqs. 15) are **not** time reversal invariant. If $\rho(t,x)$ is interpreted as charge-density then the second set of equations would be the TCP theorem. However we see from Eqs. 32 that the four group is also operative on **J**. This would be our candidate for charge-density. Notice that our particle (Eq. 19) has the highest symmetry possible.

**The Lorentz Group.The Poncare' Group:**
We have already discussed the Lorentz group. Also, it is clear that $\rho$(t-$t_0$, x-$x_0$) is a solution if $\rho$(t,x) is a solution.

**(4.6) Expressing the Equations in Generally Covariant Form:**
The equations above are covariant but only to coordinate systems moving with constant relative velocity to one another. To express them in generally covariant form we first stipulate that the ρ's are tensor densities of weight 1 (w=+1). This is because the volume element is a density of weight –1 (w = -1), and we want the integral of ρ over space to be a geometrical object[9] of weight zero. Since the divergence of a vector density is a scalar density[9] we can treat the LHS of Eqs. 15 as scalar densities. Note that not only is the divergence a scalar density, but also the LHS is the expression for the divergence in an arbitrary coordinate system.[16] All that is required for a generally covariant equation is for the RHS to be generally covariant scalar density.

To make the RHS into a scalar density we multiply the product of the two Primal-stuff vectors by the two-dimensional Levi-Civita alternating symbol $\varepsilon_{ij}$ for which w = -1. We obtain

$$\frac{\partial \rho_R^j}{\partial x^j} = \varepsilon_{ij} \rho_R^i \rho_L^j \tag{35A}$$

$$\frac{\partial \rho_L^j}{\partial x^j} = \varepsilon_{ij} \rho_L^i \rho_R^j \tag{35B}$$

The densities must be the same on each side of the equation and this fact provides a significant limitation on the possible forms of the equations (this carries over also to the case of (1+3) dimensions).

In these equations it is always true that $\rho_R^0 = \rho_R^1$ and $\rho_L^0 = -\rho_L^1$. That is to say the primal-stuff vector-density is always a null-vector-density.

Notice that these equations take no notice of the geometry of the space in which they are imbedded. They are valid no matter what the geometry is. Since these equations take no notice of geometry, matter is more primal than geometry. This is contrary to the notions of Riemann and Cliford that matter is geometry. Rather it would be more correct to suppose that the primal-stuff comprising matter in the process of propagating with speed c actually creates the space in which it then becomes imbedded. Primal-stuff is the dancer, and space is the dance hall which stuff creates in order to dance. However in (1+3)-d we have not yet been able to construct equations for **ρ(x:Ω)** without introducing the metric, so perhaps matter and geometry are of equal importance. However even in (1+3)-d we maintain that matter is distinct from geometry for the simple reason that **ρ(x;Ω)** has a two-fold infinity of values at each space point whereas no geometric quantity has a corresponding two-fold infinity of values at each space point..

Other generally covariant generalizations of Eqs. 15 are possible. One such which involves the metric is

$$\frac{\partial \rho_R^j}{\partial x^j} = g^{-1/2} g_{ij} \rho_R^i \rho_L^j \tag{36A}$$

$$\frac{\partial \rho_L{}^j}{\partial x^j} = -g^{-1/2} g_{ij} \rho_L{}^i \rho_R{}^j \tag{36B}$$

Here g is the determinant of the metric tensor for which the weight is W=2 and the metric tensor itself has W=0. These equations cannot be solved unless one knows the metric so in this case an additional equation relating primal-stuff to the metric is required.

**(4.7) Gravity in One dimension:**
Eqs. 35 are valid for an arbitrary metric; any Riemannian metric is consistent with Eqs. 35. However, we can postulate that in (1+1)-d a given connection exists. On the other hand Eqs. 36 cannot be solved without knowledge of the metric. So, in either case, we have to determine a relation between the $g_{ij}$ and the ρ**'s** .The first inclination is to follow Einstein

$$G_{ij} = \propto T_{ij} \tag{37}$$

Where **T** is a conserved tensor (yet to be) constructed from the **ρ,** and **G** is the Einstein tensor**.** But $G_{ij}$ =0 in two dimensions.[17] So we must abandon Eqs. 37 for a world of (1+1) dimensions (but not in (1+3) dimensions). Another difficulty is that the notion of particles moving along the curved geodesics of space considers particles to be point objects. But particles are not points; they are extended objects. This should not be a difficulty if the extension of the particle is small compared to the distance over which the geodesic curves.

But in our (1+1) world we do not have experiment to guide us so we shall go no further. However the procedure outlined above should serve us well in the real world of (1+3) dimensions. Perhaps it is enough for now simply to realize that in our (1+1) dimensional world the problem of gravity can be combined with particle formation (strong force) and electromagnetism (electromagnetic force) in a meaningful way.

Finally, another possibility is that even with a Lorentzian metric we could have gravity. This is because the primal-stuff equations are non-linear. For non-linear equations the sum of solutions is not a solution. To see that this is so take any two aggregates of matter for which we know the exact solutions. The sum of these solutions is not a solution. This means that there will be interaction between these two aggregates, no matter how far apart they are. This interaction might be gravity. Alternatively, it might just be a London dispersion force.

**(4.8) A Note on the (1+3) dimension Problem. Could ρ(x;Ω) be a Spinor?**
A spinor represents stuff traveling with the speed of light and also has a direction **Ω** associated with it**.** This is because a spinor times its complex conjugate is a null vector. Thus, it seems most natural to formulate spinor equations in implementing our paradigm in (1+3) dimensions. One advantage is that the auxiliary equations such as $\rho_R{}^0 = \rho_R{}^1$ and $\rho_L{}^0 = -\rho_L{}^1$ are not needed. Notice also that the RHS's of Eqs. 35 can be viewed as spinor invariants. If one replaces the LHS by spinor derivatives then a candidate equation for our world of (1+3) dimensions might be obtainable.

Penrose and Rindler[18] point out that every spinor equation can be transformed into a tensor equation and conversely, but that a simple spinor equation may be equivalent to a complicated tensor equation and/or conversely. Since as Dirac has shown the quantum mechanical wave function, at least for electrons is a 4 component spinor[19] it may be necessary when approaching an understanding of quantum mechanics to use a spinor formulation. This will be left for future work but in **Appendix D** we observe that if the Dirac spinor is interpreted as an integral over Ω of ρ(x;Ω) then perhaps equations for ρ(x;Ω) can be written which when integrated over Ω give the Dirac equations.

**(4.9) Is ρ(x;Ω) More Than Just a Complex Number? Spin and Isospin:**
We now discuss a simple generalization that introduces spin of various kinds into the formalism in a simple, self consistent way.

Suppose that there were two kinds of stuff, say **ρ and σ**. Then equations analogous to Eqs. 35 might be

$$\frac{\partial \rho_R^j}{\partial x^j} = A\varepsilon_{ij}\rho_R^i\rho_L^j + B\varepsilon_{ij}\rho_R^i\sigma_L^j + C\varepsilon_{ij}\sigma_R^i\rho_L^j + D\varepsilon_{ij}\sigma_R^i\sigma_L^j \qquad (38)$$

$$\frac{\partial \rho_L^j}{\partial x^j} = E\varepsilon_{ij}\rho_R^i\rho_L^j + F\varepsilon_{ij}\rho_R^i\sigma_L^j + G\varepsilon_{ij}\sigma_R^i\rho_L^j + H\varepsilon_{ij}\sigma_R^i\sigma_L^j$$

$$\frac{\partial \sigma_R^j}{\partial x^j} = I\varepsilon_{ij}\rho_R^i\rho_L^j + J\varepsilon_{ij}\rho_R^i\sigma_L^j + K\varepsilon_{ij}\sigma_R^i\rho_L^j + L\varepsilon_{ij}\sigma_R^i\sigma_L^j$$

$$\frac{\partial \sigma_L^j}{\partial x^j} = M\varepsilon_{ij}\rho_R^i\rho_L^j + N\varepsilon_{ij}\rho_R^i\sigma_L^j + P\varepsilon_{ij}\sigma_R^i\rho_L^j + Q\varepsilon_{ij}\sigma_R^i\sigma_L^j$$

where the coefficients would be chosen to satisfy certain symmetries and conservation laws. The quantities

$$\varepsilon_{ij}\rho_R^i\rho_R^j, \varepsilon_{ij}\rho_L^i\rho_L^j, \varepsilon_{ij}\sigma_R^i\sigma_R^j, \varepsilon_{ij}\sigma_L^i\sigma_L^j, \varepsilon_{ij}\rho_R^i\sigma_R^j, \varepsilon_{ij}\rho_L^i\sigma_L^j \qquad . \qquad (39)$$

are each zero. The latter two are zero because they are products of two null vectors lying in the same direction.

Now there are several reasons why we need to expand the concept of ρ beyond being a complex number.

First, we will want to associate with particles a wavelength that is proportional to particle velocity in order to satisfy the experimental observation that particles have such a wavelength. Being extended objects our particles have the a priori possibility of having such spatial/temporal variation. However the equations so far derived in this paper do not seem to allow such variation. The equations are consistent with any initial conditions and they describe how the distributions evolve in space and time. What we are saying is that we have not yet found any periodically varying initial distribution that maintains itself in existence at a given place (or is moving at a constant velocity), and therefore represents a particle with wavelength. In Reference 2 we assumed that ρ was a 2x2 matrix and this introduced spin space which allowed the solutions to the equations to have spatial and temporal oscillations in spin space, thus giving rise to particle wavelength. Further this wavelength was inversely proportional to particle velocity in accordance with experiment.[2] Particle wavelength is a quantum mechanical necessity. See **Appendix D** for a discussion of some QM implications of our paradigm.

Second, the complexities of our real world may well require that ρ being merely a complex number is not enough. This can be handled by making ρ a matrix. So effectively there is a tensor equation for each element of the matrix. This allows us to to introduce spin into the equations and perhaps in this way the weak force is incorporated.

**(4.10) A (1+1) Dimensional World in Which ρ is an nxn Matrix:**
Products of R-R matrices and L-L matrices do not exist; they are equal to zero (see Eq. 39). So, if we want to generalize Eqs. 35 to be matrix equations then the matrices on the RHS of Eqs. 35 can be only R-L and L-R matrices

The suggested equations[3] in a (1+3) dimensional world when applied to (1+1) dimensions become.

$$\frac{\partial \rho_R^j}{\partial x^j} = \varepsilon_{ij}(\rho_R^i \rho_L^j + \rho_L^i \rho_R^j) \tag{40A}$$

$$\frac{\partial \rho_L^j}{\partial x^j} = -\varepsilon_{ij}(\rho_R^i \rho_L^j + \rho_L^i \rho_R^j) \tag{40B}$$

where the ρ are now nxn matrices, each element of which is a vector density. These equations are covariant to general relativity. Realizing that $\rho_R^0 = \rho_R^1 = \rho_R$ and $\rho_L^0 = -\rho_L^1 = \rho_L$ the equations can be written as

$$\frac{\partial \rho_R}{\partial t} + \frac{\partial \rho_R}{\partial x} = -2[\rho_R, \rho_L] \tag{41A}$$

$$\frac{\partial \rho_L}{\partial t} - \frac{\partial \rho_L}{\partial x} = +2[\rho_R, \rho_L] \tag{41B}$$

[,] being the commutator ([A,B] = -[B,A]).

**(4.10.1) The Matrix is 3x3:**
If we now specify ρ to be a 3x3 matrix we can show the possibility of classifying particles in an eight-fold way. This is suggestive of the (1+3) dimensional world.

If we now expand the ρ's in terms of the 8 generators $\lambda_k$ of SU3, plus the unit matrix $\lambda_0$ (any 3x3 matrix can be so expanded) we obtain

$$\rho_R = \sum_{0}^{8} A_k \lambda_k \tag{42A}$$

$$\rho_L = \sum_0^8 B_\ell \lambda_\ell \tag{42B}$$

Placing Eqs. 42 into Eqs. 41 we obtain

$$\frac{\partial(\sum_0^8 A_k \lambda_k)}{\partial t} + \frac{\partial(\sum_0^8 A_k \lambda_k)}{\partial x} + 2\sum_{0,0}^{8,8} A_k B_l [\lambda_k, \lambda_\ell] = 0 \tag{43A}$$

$$\frac{\partial(\sum_0^8 B_l \lambda_l)}{\partial t} - \frac{\partial(\sum_0^8 B_l \lambda_l)}{\partial x} - 2\sum_{0,0}^{8,8} A_k B_l [\lambda_k, \lambda_\ell] = 0 \tag{43B}$$

Using the relation $[\lambda_k, \lambda_\ell] = 2i C_{k\ell m} \lambda_m$ where $C_{k\ell m}$ are the structure constants (no sum over m) and the fact that

$$\sum_0^8 D_m \lambda_m = 0 \tag{44}$$

implies that each of the nine $D_j$ are zero, we obtain nine equations for the A's and nine for the B's. We obtain

$$\frac{\partial A_m}{\partial t} + \frac{\partial A_m}{\partial x} = -4i \sum_{0,0}^{8,8} A_k B_l C_{klm} \tag{45A}$$

$$\frac{\partial B_m}{\partial t} - \frac{\partial B_m}{\partial x} = +4i \sum_{0,0}^{8,8} A_k B_l C_{klm} \tag{45B}$$

Using the standard values for the structure constants we obtain

$$\frac{\partial A_0}{\partial t} + \frac{\partial A_0}{\partial x} = 0 \tag{46A}$$

$$\frac{i}{2}\left(\frac{\partial A_1}{\partial t} + \frac{\partial A_1}{\partial x}\right) = 2(A_2 B_3 - B_2 A_3) + (A_4 B_7 - B_4 A_7) - (A_5 B_6 - B_5 A_6)$$

$$\frac{i}{2}\left(\frac{\partial A_2}{\partial t} + \frac{\partial A_2}{\partial x}\right) = 2(A_3 B_1 - B_3 A_1) + (A_4 B_6 - B_4 A_6) + (A_5 B_7 - B_5 A_7)$$

$$\frac{i}{2}\left(\frac{\partial A_3}{\partial t} + \frac{\partial A_3}{\partial x}\right) = 2(A_1 B_2 - B_1 A_2) + (A_4 B_5 - B_4 A_5) - (A_6 B_7 - B_6 A_7)$$

$$\frac{i}{2}(\frac{\partial A_4}{\partial t}+\frac{\partial A_4}{\partial x}) = (A_7 B_1 - B_7 A_1) + (A_6 B_2 - B_6 A_2) + (A_5 B_3 - B_5 A_3) + \sqrt{3}(A_5 B_8 - B_5 A_8)$$

$$\frac{i}{2}(\frac{\partial A_5}{\partial t}+\frac{\partial A_5}{\partial x}) = -(A_6 B_1 - B_1 A_6) + (A_7 B_2 - B_7 A_2) + (A_3 B_4 - B_3 A_4) + \sqrt{3}(A_8 B_4 - B_8 A_4)$$

$$\frac{i}{2}(\frac{\partial A_6}{\partial t}+\frac{\partial A_6}{\partial x}) = -(A_1 B_5 - B_1 A_5) + (A_2 B_4 - B_2 A_4) - (A_7 B_3 - B_7 A_3) + \sqrt{3}(A_7 B_8 - B_7 A_8)$$

$$\frac{i}{2}(\frac{\partial A_7}{\partial t}+\frac{\partial A_7}{\partial x}) = (A_1 B_4 - B_1 A_4) + (A_2 B_5 - B_2 A_5) - (A_3 B_6 - B_3 A_6) + \sqrt{3}(A_8 B_6 - B_8 A_6)$$

$$\frac{i}{2}(\frac{\partial A_8}{\partial t}+\frac{\partial A_8}{\partial x}) = \sqrt{3}(A_4 B_5 - B_4 A_5) + \sqrt{3}(A_6 B_7 - B_6 A_7)$$

The equations for the nine Bs are similar. The LHS reads $\frac{i}{2}(\frac{\partial B_j}{\partial t} - \frac{\partial B_j}{\partial x})$ and the RHS is the negative of the corresponding RHS above. The equation for $B_1$ reads

$$-\frac{i}{2}(\frac{\partial B_1}{\partial t} - \frac{\partial B_1}{\partial x}) = 2(A_2 B_3 - B_2 A_3) + (A_4 B_7 - B_4 A_7) - (A_5 B_6 - B_5 A_6) \qquad (46B)$$

And so on.

These equations are interesting because they suggest the existence of various kinds of particles. For example if $A_1, A_2, A_3$ ; $B_1, B_2, B_3$, are non-zero but all other A's and B's are zero then the equations would evolve in time and space with only those A's and B's indexed by 1, 2, 3 being non-zero. All the other A's and B's would not appear. There are a total of 8 classes of particles. Their indices are

123;  147;  156;  246;  257;  3458;  3678;  12345678 (47)

Because our equations give the dynamics of how these particles move in space and time (in our one-dimensional world) we are able to say how stuff described by pairs of non-zero A's and B's evolve (the other A's and B's are initially presumed to be zero). Examination of Eqs. 46 shows that

1,2 and 2,3 and 3,1 each evolve to 123. (48)

1,4 and 1,7 and 47 each evolve to 147

1,5 and 5,6 and 1.6 each evolve to 156

2,4 and 2,6 and 4,6 each evolve to 246

2,5 and 2,7 and 5,7 each evolve to 257

3,4 and 3,5 and 4,5 and 4,8 and 5,8 each evolve to 3458

3,6 and 3,7 and 6,7 and 6,8 and 7,8 each evolve to 3678.

1,8 and 2,8 and 3,8 display no coupling. Thus a pair either generates other A's and B's or there is no coupling.

A triplet either generates one of the 7 classes enumerated above or it generates all 8 A's and B's. A singlet just stays single and corresponds to propagation with speed c

$A_0$ and $B_0$ never couple to any of the other 8 A's and B's (except possibly through the metric). Their stuff is ghost stuff. In our real world of (1+3) dimensions $A_0$, $B_0$ might be dark matter/dark energy. Notice that there could not be dark matter/dark energy particles.

**(4.10.2) Additional Symmetries:**
In addition to the scaling group, the four group, and the Lorentz-Poincare group mentioned previously, there are 4 groups arising from the fact that the ρ's are now matrices. They are the transpose group -$\tilde{\rho}$, ( -( $\tilde{-\tilde{\rho}}$ )= ρ ); the complex conjugate $\rho^*$, ( $(\rho^*)^*$ = ρ ); and the adjoint (conjugate transpose) group -$\rho^\dagger$, (-(-$\rho^\dagger$)$^\dagger$ = $\rho^\dagger$ ). Also, M**ρ**M$^{-1}$ where M is any non-singular constant matrix is a group operation. Group operations can generate previously unknown solutions from known solutions and sometimes new particles from known particles.
.
**(4.10.3) Conservation Laws, Invariants and Such:**
Without conservation laws and invariants our world would be a chaotic one; there would be no stabilities. Fortunately our equations display both.

In our real world electric charge is both a conserved quantity and an invariant quantity. The conservation law is well known (the divergence of charge-current is zero), but we observe that it is also an invariant quantity. This is because the integral of charge density over space is an invariant. Thus no matter with what velocity we approach a charge distribution the total charge of that distribution is independent of the velocity of approach. Presumably, spin since it also does not depend on velocity of approach is a vector density also. Energy however is a conserved quantity, but it is not an invariant quantity. The conservation is expressed as the divergence of the energy tensor equaling zero, but because it is a tensor it is a function of the velocity with which we approach the energy distribution.

Further, in studies of the Kurtewig-Devries equation[20] we see that one equation can have many conservation laws associated with it[13] (see also **Section 4.2.1.** We therefore must be open to the possibility, indeed expect, that our equations in both (1+1)-d and (1+3)-d can have multiple conservation laws.

The conservation law obtained by adding Eqs. 41A and 41B is a conservation law on each element of the matrix. This conservation law carries through to each of the particles classified in Eq. 47. Thus, for example, for the particles labeled 123 we have

$$\frac{\partial Q^{\mu}_{123}}{\partial x^{\mu}} = o \tag{49}$$

where $Q^{\mu}_{123} = Q^{\mu}_1 + Q^{\mu}_2 + Q^{\mu}_3$, $(Q^0_1, Q^1_1) = (A_1 + B_1, A_1 - B_1)$

Just as in Eqs. 32 we can write

$$\Box Q^{\mu}_{123} = J^{\mu}_{123} \quad , \quad \frac{\partial J^{\mu}_{123}}{\partial x^{\mu}} = 0 \tag{50}$$

The conserved 2-current (conserved 4-current in 4-d) if non-zero would correspond to charge-current. The 123 equations could have multiple solutions, some with charge and some without charge.

Notice that because **Q** is a vector-density Eq. 49 is generally covariant.[16]

Notice that $A^{\mu}_0$ also obeys a conservation law. Thus there are 9 in total.

Other conservation laws arise from the matrix character of the equations. If we multiply the jth equation of the set 46A by $A_j$ and then sum over j we obtain

$$\frac{\partial \sum (A_j)^2}{\partial t} + \frac{\partial \sum (A_j)^2}{\partial x} = 0 \tag{51A}$$

Similarly, we have

$$\frac{\partial \sum (B_j)^2}{\partial t} - \frac{\partial \sum (B_j)^2}{\partial x} = 0 \tag{51B}$$

where the sum is over those indices describing the particle in question. Thus for particle 156 of the set (47) the sum would be over j =1, 5, 6 since only these A's are non-zero. Apparently these quantities are Casmir invariants.[21] If so there should also be a cubic invariant.

Further, any function of $\sum (A_j)^2$ and/or $\sum (B_j)^2$ obeys conservation laws.

Whenever we have a group we have group invariants. Thus all in all we have plenty of possibilities to describe mass, charge, and spin. It is perhaps best to stop here in the presentation of our one dimensional world and wait for the real world of 3 dimensions to guide us in our selection and identification of the conserved quantities of physics.

**(5.0) Summary and Conclusions:**
We began by deriving the Lorentz transformation (LT) without assuming the existence of Maxwell's equations, or that the speed of light is a constant, or even that light exists. A quantity c with the dimensions of velocity appears in our derivation of the LT. It is a property of the LT that an object moving with speed c in one coordinate system moves with speed c in every coordinate system. The experimental fact that light has this exact property allows us to set c=2.998x10$^8$ m/sec. This fact becomes a great confidence builder. It suggests that pure thought is adequate to the task of deriving the laws of physics and establishes the LT as the prime epistemological reality upon which to build the edifice of physical law.

Using Ockham's razor and the requirement that the equations of physics should be Lorentz covariant led us to postulate that there exists in all of physics only one speed. That speed is the speed of light, c. This leads us to construct a field **ρ(**t,x,y,z**;Ω)** in which the quantity **ρ** at every place **x** and in every direction **Ω** (at every place **x**) travels at the speed of light**.** See Fig. 1. We call this field primal-stuff. We make the bold hypothesis that the primal-stuff field is the only thing that exists and that therefore all the laws of physics are implicit in **ρ(**t,x,y,z**;Ω).** This paradigm allows us to observe that Riemannian geometry and Weyl geometry are one and the same because transference of length is achieved by **ρ** moving with speed c in every direction **Ω** at every point **x**. See **Appendix A**.

To make the above ideas clearer and familiar, we, in the remainder of the paper, constructed equations for a world in which time and only one space direction occur.

First, we devised equations (the LT is valid globally) in which primal-stuff traveling to the right scatters and is scattered by primal-stuff traveling to the left in such a manner that the equations are covariant to the Lorentz transform (See Eqs. 15). Particles, which are necessarily extended objects, are obtained which can be stationary even though their constituent parts $\rho_R$ and $\rho_L$ are always traveling at the speed of light. And because the equations are Lorentz covariant the particles can move with any constant speed v (-c<v<c). Specific examples are given of one-particle and two-particle states. These particles are quantized by virtue of the non-linearity of the equations. We emphasize that the equations predict the existence of the particles; they are not assumed.

The general solution to the equations is obtained in terms of two arbitrary functions g(x+t) and f(x-t); see Eqs. 23.

Derivation of the Doppler effect, the concept of mass and $E=mc^2$, conservation laws, symmetries, and the creation and annihilation interaction of the particles of our (1+1) dimensional world allow us to conclude that the strong and electromagnetic forces are embodied in our equations.

To incorporate gravity we generalize the equations so they are valid in every arbitrary coordinate system (Eqs. 35, 36). A complete description of our (1+1)-d world now requires an additional equation relating primal-stuff to the metric. See **Sections 4.6** and **4.7**. With such a selection we will have unified the strong force, electromagnetism and gravity.

Our equations up to this point seem to lack spin (and hyperspin). To incorporate spin we generalize **ρ(**t,x,y,z**;Ω)** so that it is a matrix. We discuss the case where **ρ(**t,x,y,z**;Ω)** is 3x3. This leads to an eight-fold way in which eight classes of particles exist corresponding to eight primal-stuff fields. The ninth field cannot support particles and does not interact with the other fields, except possibly through the metric. Might this be dark matter/energy?

Symmetries and conservation laws are obtained and discussed.

In **Appendix B** we develop Lagrangians which when minimized lead to previously obtained equations.

In **Appendix C** we offer a candidate equation for our real world of (1+3)-d.

**Appendix D** discusses the difficult problem of incorporating quantum mechanics into the new paradigm.

Thus, in this paper we have given a somewhat detailed treatment of our paradigm for the case of (1+1) dimensions, although much more needs to be done. Hopefully an intuition into the more difficult real world case of (1+3) dimensions has been developed.

   It is hoped that these ideas provide a viable alternative to string theory. Essentially our position is that we live in a world of 3 spatial dimensions and time and that the particles of this world are extended objects in these three dimensions. Accordingly particles are not points, strings, or closed surfaces, nor are they m-dimensional objects embedded in n-dimensional space where 3<m<n. The author believes that the possible paradigms applicable to our real world of (1+3) dimensions should be exhausted before considering the higher dimensional possibilities of string theory. To quote J. Willard Gibbs "One of the principle objects of theoretical research is to find the point of view from which the subject appears in its greatest simplicity".[22]

**Apendix A, Does Our Paradigm Dictate Geometry?**

In the text we assumed the Riemannian metric,

$$ds^2 = g_{ij}dx^i dx^j \qquad (A1)$$

but geometry can be more general. Weyl showed[8] that in order to account for the transference of length, $\ell$, from point to point within a manifold a new function

$$d\varphi = \varphi_j dx^j \qquad (A2)$$

must be defined, in addition to the metric. The transference of length $\ell$ is made through the relation $d\ell = -\ell d\varphi$. Although geometers accept this idea, for physicists there are 3 problems. First there are many different possible choices for $\varphi_j$ so why should we prefer one choice over another? Second, Weyl was unable to show that the idea had the richness necessary to explain the richness of physics. Third, Weyl identified $\varphi_j$ as the electromagnetic 4-vector $A_j$ and this conflicts with certain laws of spectroscopy, as Einstein and others showed. Perhaps we could satisfy the demands of both geometers and physicists by suggesting that light itself is the proper measure for transferring length.

If we accept that the only existing physical reality is primal-stuff then it is primal-stuff alone that must be employed to construct $\varphi_j$. Now consider that the primal-stuff 4-vector always travels with speed c, and c is a null vector so that we can write

$$c_j dx^j = 0, \qquad (A3)$$

where $c_j = (c_0, c_1, c_2, c_3)$ is the covariant velocity 4-vector for the speed of light, and importantly $dx^j$ is in the same direction as $c_j$. Identification of $c_j$ with $\varphi_j$ immediately gives $d\varphi = 0$ because $c_j dx^j = 0$. This establishes transference of length but only along this one direction. But the same argument can be used for every direction $\boldsymbol{\Omega}$ and for every point **x** of the manifold. Thus $d\varphi = 0$ and therefore $d\ell = -\ell d\varphi = 0$. The value of $\lambda$ itself is now a constant throughout all space and is chosen by convention.

Thus with one stroke our paradigm collapses Weyl geometry into Riemannian geometry as they apply to our real world and they become one and the same. Note that $c_j (t,x,y,z,\boldsymbol{\Omega}) dx^j = 0$ is valid for every point and for every direction $\boldsymbol{\Omega}$. Some might prefer to write it as $c_j (t,x,y,z,\boldsymbol{\Omega}) dx^j d\boldsymbol{\Omega} = 0$. since this would remind us that in (1+3)-space the appropriate 4-vector density is $\rho^j d\boldsymbol{\Omega}$.

With our paradigm the view that physics is geometry dies; rather robust matter being composed of primal-stuff always traveling with speed c, creates the Riemannian geometry from the void into which it expands. A more poetic description might be that primal-stuff, in order to do the dance of life, creates the ballroom in which to dance.

**Appendix B, Equivalence of our Equations to a Variational Principle:**

In reference 1 we showed that Eqs. 15 are equivalent to

$$\frac{\partial^2 F}{\partial t^2} - \frac{\partial^2 F}{\partial x^2} = \left(\frac{\partial F}{\partial t}\right)^2 - \left(\frac{\partial F}{\partial x}\right)^2 \tag{B1}$$

where

$$\frac{\partial F}{\partial x} = Q_0 \quad , \quad \frac{\partial F}{\partial t} = -Q_1 \tag{B2}$$

and

$$Q_0 = \rho_R + \rho_L \quad , \quad Q_1 = \rho_R - \rho_L \tag{B3}$$

But Eq. B1 can be obtained from

$$\int e^{-F} \left( \left(\frac{\partial F}{\partial t}\right)^2 - \left(\frac{\partial F}{\partial x}\right)^2 \right) dt dx = extremum \tag{B4}$$

This is true as long as F is viewed as being pure real. But in the derivation and the solutions (see text) F is allowed to be complex, so the quantity F in Eq. B1 is allowed to have both real and imaginary parts. In order to derive the **complex** equation from an extremum principle we need to find an invariant "Lagrangian" that is a function of both $F_R$ and $F_I$ where they are the real and imaginary parts of F ( $F = F_R + iF_I$).
  Similar considerations apply to Eq. 15 of Reference 2. (There is a typo in Eq. 15. The RHS should be multiplied by i =√(-1)). Eq. 15 as it reads in Reference 2 is for **G'** =i**G**. **G** is the 3 component spin vector. The equation leading to this is

$$\int \left( \left(\frac{\partial G'}{\partial t}\right)^2 - \left(\frac{\partial G'}{\partial x}\right)^2 + \left(\frac{4}{3}\right) G' \cdot \left(\frac{\partial G'}{\partial x}\right) \times \left(\frac{\partial G'}{\partial t}\right) \right) dt dx = extremum \tag{B5}$$

An extremum principle that includes both real and imaginary parts would be a function of 6 dependent variables. We have not yet investigated the problem of obtaining Lagrangians whose resulting equations allow for complex and hypercomplex numbers. Noether's theorem[23] can be used to derive conservation laws from invariant Lagrangians, so it would put us on familiar ground if such Lagrangians existed.

**Appendix C, World of (1+3)-Dimensions:**

In Reference 3 we suggested a Lorentz covariant equation for the (1+3)-d primal-stuff vector. Here we generalize to general covariance. At each point in space-time we imagine ourselves to be riding along the null vector density in the direction $\mathbf{\Omega}$ with the speed of light. Consider the following equation.

$$\frac{\partial(\rho^\mu(\Omega)d\Omega)}{\partial x^\mu} = g^{-1/2} g_{\mu\nu}[\rho^\mu(\Omega)d\Omega, \int \rho^\nu(\Omega_1)d\Omega_1] \tag{C1}$$

All of the two-fold infinities of direction $\mathbf{\Omega}$ are allowed at each point. See Fig. 1. For the case of Lorentz covariance the LHS of Eq. C1 is the comoving derivative (also called material or substantial derivative). But more generally the LHS is the four-divergence of a vector density and it is a scalar density.[16] It is also covariant to an arbitrary coordinate transformation.[16] We choose $\rho(x;\Omega)d\Omega$ to be a **null-four-vector-density-matrix**. A **density** because we want the LHS of Eq. C1 to be a scalar density; **null** because according to our paradigm the only thing that exists is primal-stuff traveling always with the speed of light; a (1+3) **four-vector** to maintain consistency with the only geometry possible which is locally, at least, Lorentzian; an nxn **matrix** to insure the possibility of a particle zoo. $\rho(x;\Omega)d\Omega$ is the vector density and $d\Omega$ is a solid angle. The brackets are the commutator [A,B]=-[B,A]. Notice that when we integrate over $\mathbf{\Omega}$ we obtain a conservation law on each of the matrix elements.

$$\frac{\partial(\int \rho^\mu(\Omega)d\Omega)}{\partial x^\mu} = 0 \tag{C2}$$

Generally the RHS is to be constructed from all possible combinations of the stuff vector density which is $\rho(x;\Omega)d\Omega$; the following numerical tensors, the Levi-Civita alternating symbols $\varepsilon_{\mu\nu\lambda\sigma}, \varepsilon^{\mu\nu\lambda\sigma}$, the Kronecker deltas $\delta^\mu_\alpha, \delta^{\mu\nu}_{\alpha\beta}, \delta^{\mu\nu\lambda}_{\alpha\beta\delta}, \delta^{\mu\nu\lambda\sigma}_{\alpha\beta\delta\varepsilon}$, and the metric tensor $g_{\mu\nu}$ since according to our view they are the only things that exist. The RHS of Eq. C1 is one possibility. For the determinant of the metric g the weight is 2 (W=2); for $\rho(x;\Omega)d\Omega$, W=1; and for $g_{\mu\nu}$, W=0; so the RHS is a scalar density. We have not been able to construct other candidates for the RHS.

Eq. C1 is then generally covariant. If it admits only of the Lorentz metric globally as well as locally then it would be the total reality. Otherwise Eq. C1 plus the Einstein equation,

$$G_{\mu\nu} + g_{\mu\nu}\Lambda = -8\pi G T_{\mu\nu} \tag{C3}$$

taken together are the candidate equations for our real world. Here $T_{\mu\nu}$ is a yet to be determined divergence-less tensor constructed from the primal-stuff field and its derivatives.

If we accept these equations then the notion of Riemann and Clifford that matter is essentially geometry dies. This is because geometry cannot specify the primal-stuff field which has at each space time point a two-fold infinity of possible values arising from the two-fold infinity of directions $\mathbf{\Omega}$.

**Appendix D, Deriving Quantum Mechanics:**

**(D1)** We always have the option of viewing the work of this paper as a statement of classical physics and quantizing it just as was done previously, in the early years, to obtain quantum mechanics. In this way a quantum mechanical version of the theory presented here may be possible. However we first should see if our paradigm allows QM to be derived directly.

**(D2)** Mathematicians familiar with the inverse problem know that given an energy level structure, and the scattering coefficients, one can derive a Sturm-Liouville equation and a potential energy which when put into the Sturm-Liouville equation reproduces the energy level structure.[24] This means that if our (1+1) dimensional equations display an energy level structure then we should be able to derive a time independent Schrodinger equation (time independent Schrodinger equation equals Sturm-Liouville equation) and a potential energy which reproduces the energy level structure. The Schrödinger equation will treat the particles as points even though we know they are extended objects. To say it in another way; the Schrödinger equation assumes particles to be points but this is no reason for believing the particles are points even though exact predictions of energy levels are obtained. An implication of the above observations is that a part of QM is mere bookkeeping (in complete accordance with the philosophy of positivism which was prevalent at the time). This is because solving the inverse problem always results in a Schrodinger equation; only the potential energy changes. To say it in another way; if the experimentalists suddenly confessed to giving the theorists the wrong energy level structure all these years it would be no problem for the theorist to derive the new energy level structure. Only the potential energy would change in the Schrodinger equation. In a real sense the only meaningful question is why is the potential energy in 3-d inverse in r. Thus we can argue that doing quantum mechanics via the time independent Schrodinger equation is mere bookkeeping. Second Quantization can also be understood as bookkeeping because it can accommodate any energy level structure no matter what it is.

However the Schrodinger equation is only one aspect of QM. Also, we have not yet found an energy level structure in our treatment. We leave this to future work. This is not to say that QM is not a great advance. It is. And in fact our view of physical reality must either explain QM or at least be consistent with it.

**(D3)** We must also make an attempt to discuss the other aspects of QM. One such is the association of a wavelength to a particle which is in inverse relation to the velocity of the particle. Because the particles are extended objects it is easy to envisage how a particle can have a wavelength that is inverse to its velocity. Suppose that our one-dimensional particle has a spatial oscillation. Such a particle (imagine an electron) will have maxima and minima in space and time. Then, if we throw the electron against a crystal with velocity 2V the maxima will hit atoms in the crystal with twice the frequency than if we threw the particle at the crystal with the velocity V. This explains why the electron wavelength is inverse to the velocity. The relativistic formula was derived in Reference 2.[2] Also, if we threw the crystal at a stationary electron we would still have to high accuracy the same situation because of Lorenz covariance. The wavelengths of the atoms would, to a high approximation, not be relevant to our calculation of the scattering laws.

To say that QM is explained we would have to explain all of the paradoxical behaviors. We have not done this; but consider the following. A particle may really be quite a big object so that there is a real sense in which it can be (parts of it can be) in many places at once. Coupling this with the fact that its constituent parts are moving with the speed of light it may be quite possible that in the collapsing of a wave packet it can have its center move quickly to any place within its location as described by its quantum mechanical wave function.

The experiments of Aspect and others seem to show that QM entities can travel faster than the speed of light. They are rather convincing. However they assume that particles are point objects. For their experiments to be completely convincing their proofs must allow for the particles to be extended objects at their time of formation.

**(D4)** Perhaps the most insightful investigation into Quantum Mechanics (QM) was by Dirac who showed that The wave function of electrons could be viewed as a 4-component spinor[19] which is a bispinor (the sum of two two-component spinors which we call $\varphi_1$ and $\varphi_2$. Now a two component spinor can be interpreted as representing stuff traveling with the speed of light in a given direction $\Omega$. That this is so becomes evident when it is realized that a spinor times its complex conjugate is a null vector and the direction that the stuff is travelling in with the speed of light is that of the null vector. Now the sum of spinors is a spinor. This must mean that if you are given a spinor, the direction and magnitude that it represents may be that of a sum of spinors. Let us write

$$\varphi_1 = \int \sigma_1(x:\Omega)d\Omega$$
$$\varphi_2 = \int \sigma_2(x:\Omega)d\Omega$$
(D1)

Where we call the sigmas primal-spinors. Now the usual interpretation of zitterbewegung is that it is a jittering of wave function caused by the periodic feeding into and out of the two spinors $\varphi_1$, $\varphi_2$ forming the bispinor thereby allowing the electron to travel at a speed less than that of light even though the individual two spinors are jittering at the speed of light. Just as Dirac went beneath the Klein-Gordon equation by effectively taking a square root, so too can we go beneath the Dirac equation by writing equations for $\sigma_1$(t,x,y,z;$\Omega$) and $\sigma_2$(t,x,y,z:$\Omega$) which when integrated over $\Omega$ gives the Dirac equation?

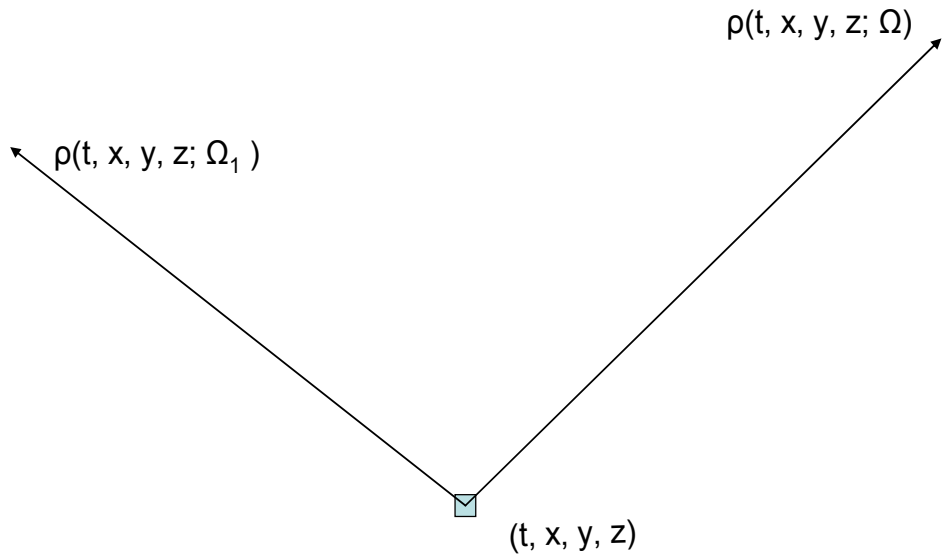

Figure 1. The amount of primal-stuff **ρ(**t,x,y,z;**Ω)** at (t,x,y,z) traveling in direction **Ω,** always with the speed of light, changes by being scattered by other primal-stuff **ρ(**t,x,y,z;**Ω_j)** at the same space-time point. All **Ω_j** and **Ω** are allowed simultaneously. The game is to choose the interaction law so that certain configurations of field persist in place even though stuff is always traveling with the speed of light. Such configurations are called particles. In this paper, for a world of (1+1) dimensions, we show that quantized particle formation, electromagnetism, gravitation, and hyperspin can all be treated in a unified self- consistent manner.